\documentclass[11pt,superscriptaddress,aps,prd,preprint]{revtex4}
\usepackage{amsmath}

\makeatletter
\usepackage[T1]{fontenc}
\usepackage{amsmath}
\usepackage{amssymb}
\usepackage{graphicx}

\usepackage{slashed}

\DeclareSymbolFont{extraitalic}      {U}{zavm}{m}{it}
\DeclareMathSymbol{\Qoppa}{\mathord}{extraitalic}{161}
\DeclareMathSymbol{\qoppa}{\mathord}{extraitalic}{162}
\DeclareMathSymbol{\Stigma}{\mathord}{extraitalic}{167}
\DeclareMathSymbol{\Sampi}{\mathord}{extraitalic}{165}
\DeclareMathSymbol{\sampi}{\mathord}{extraitalic}{166}
\DeclareMathSymbol{\stigma}{\mathord}{extraitalic}{168}

\newcommand{\bea}{\begin{eqnarray}}
\newcommand{\eea}{\end{eqnarray}}

\begin{document}

\title{Lagrangian formalism for Rastall theory of gravity and G\"{o}del-type universe}

\author{W. A. G.  De Moraes}\email[]{welton@fisica.ufmt.br }
\affiliation{Instituto de F\'{\i}sica, Universidade Federal de Mato Grosso,\\
78060-900, Cuiab\'{a}, Mato Grosso, Brazil}

\author{A. F. Santos}\email[]{alesandroferreira@fisica.ufmt.br}
\affiliation{Instituto de F\'{\i}sica, Universidade Federal de Mato Grosso,\\
78060-900, Cuiab\'{a}, Mato Grosso, Brazil}

\begin{abstract}

In the Rastall gravity a non-minimal coupling between geometry and matter fields is considered. Then the usual energy-momentum tensor conservation law is not valid. Here a Lagrangian formalism is proposed to the Rastall theory of gravity. The G\"{o}del-type universe is studied in this gravitational model. Then it is studied whether this theory permits causality violation. The field equations do not exclude solutions with a breakdown of causality for a perfect fluid as matter content. In this case, an expression for the critical radius (beyond which the causality is violated) is determined. In addition, for a combination between perfect fluid and scalar field as matter content the theory accommodates causal G\"{o}del-type solution.

\end{abstract}

\maketitle

\section{Introduction}

The General Relativity (GR) is a classical theory of gravitation. Although GR is successful tested there are various observational data that it does not explain, such as the current accelerating phase of the universe expansion, the dark matter problem, among others. In addition there is not a quantum gravity theory completely consistent. These problems lead to modified theories of gravity (for a review see \cite{review}). In GR and in most of these modified theories the covariant conservation of the energy-momentum tensor is a well known ingredient which via the Noether symmetry theorem leads to the conservation of some physical quantities. However, some modified theories have proposed that the condition of covariant energy-momentum tensor conservation can be relaxed. From these ideas, P. Rastall in 1972 constructed a new theory of gravity where the usual conservation law is questioned \cite{Rastall, Rastall1}.

In Rastall theory the usual energy-momentum tensor conservation law is not satisfied. In this gravitational model a non-minimal coupling of matter fields to geometry is considered where the divergence of $T_{\mu\nu}$ is proportional to the gradient of the Ricci scalar, i.e., $\nabla^\mu T_{\mu\nu}=\nabla_\nu R$. However, the usual conservation law is recovered in the flat space-time. This proposal has as a strong argument the fact that the usual conservation law on energy-momentum tensor is tested only in the flat space-time or in a gravitational weak field limit. Several studies have been developed with this theory. For example, a phenomenological way for distinguishing features of quantum effects in gravitational systems has been reproduced \cite{Koi, Min}, a classical formulation for the particle creation in cosmology has been considered \cite{Batista}, many interesting features of the cosmological models have been investigated \cite{Fabris}, some rotating and non-rotating black hole solutions have been studied \cite{Hey, Hey1, Kumar, Ma}, application of the Newtonian limit to the theory has been analyzed \cite{Mor}, the traversable asymptotically flat wormhole solutions has been investigated \cite{Mor1}, a generalization of the Rastall theory has been realized to discuss the cosmic accelerated expansion \cite{Mor2}, a gravitational collapse of a homogeneous perfect fluid has been studied \cite{Ziaie}, among other. Comparisons between the Rastall theory of gravity and GR have been done. Visser \cite{Visser} has concluded that these two theories are equivalents, while Darabi et al. \cite{Darabi} suggested that the two gravitational theories are not equivalent and added that Rastall theory is more general than GR and it prepared to investigate the challenges of cosmological observations and may indicate a way to quantum gravity.

Although the Rastall gravity theory has been intensively studied recently there is a serious criticism related to this model. The criticism corresponds to the fact that this theory of gravitation is not a Lagrangian based theory (at least in the context of the Riemaniann geometry). However there are some attempts to obtain the Rastall field equations from a Lagrangian formalism \cite{Smalley, Fabris1, Fabris2}. The violation of the classical conservation law of the energy-momentum tensor is not an exclusivity of the Rastall gravitational theory, there are other models that also reported this characteristic, such as $f(R,T)$ gravity \cite{Harko}, $f(R,{\cal L}_m)$ gravity \cite{Harko1, Harko2}, among others. Here these two gravitational models is used to construct a possible Lagrangian formalism to the Rastall gravity. Then the causality violation issue is studied in the Rastall theory. For this analysis the G\"{o}del-type universe is considered.

GR and some modified gravity theories allow exact solutions that lead to time travel. These solutions are well-known to encompass violations of the most intuitive notions of chronological order and causality. Several space-time models share this property and present the so-called Closed Time-like Curves (CTCs) that allow causality paradoxes and time travel. The best-known solutions containing CTCs in GR are G\"{o}del \cite{Godel}, Van-Stockum \cite{Van, Van1}, Kerr black hole \cite{Kerr, Kerr1}, cosmic strings \cite{string, string1, string2}, among others. In addition, a generalization of the G\"{o}del solution, denominated by G\"{o}del-type metrics, has been constructed \cite{Reboucas}. These metrics permit to eliminate the CTCs for determined values of their parameters.  Several studies have been developed with respect to the causality aspects of the G\"{o}del-type metrics in modified theories of gravity. The existence of completely causal solutions within these theories have been found. Such study has been realized in $f(R)$ gravity \cite{Reboucas1, Reboucas2}, $f(R,T)$ gravity \cite{AFS, AFS1}, Brans-Dicke gravity \cite{Aguirre}, Chern-Simons gravity \cite{AFS2}, Horava-Lifshitz gravity \cite{Yu}, among others. Our aim in this paper is treating the G\"{o}del-type metrics in Rastall theory and discuss the causality issue.

This paper is organized as follows. In section II, some details about the Rastall theory of gravity are presented. In section III, a Lagrangian formulation for Rastall theory is discussed. The $f(R,T)$ and $f(R, {\cal L}_m)$ modified gravity theories are considered. Then the Rastall theory is interpreted as a particular case of these theories of gravity. In sections IV, a brief introduction to the G\"{o}del-type universe is given. In section V, the G\"{o}del-type solution in Rastall theory is studied. Causal and non-causal solutions are analyzed. In section VI, some concluding remarks are discussed. In this paper, we adopt units where $G = c = 1$ unless otherwise stated.

\section{Rastall theory of gravity}

Here a brief introduction to Rastall gravity theory is presented. In general relativity a conservation law that leads to a divergence-free energy-momentum tensor  has been adopted, i.e.,
\bea
 \nabla^{\nu}T_{\mu\nu} = 0,
\eea
where $\nabla^{\nu}$ is the covariant derivative. In contrast to this idea, P. Rastall proposed a different conservation law which generalizes the general relativity ideas \cite{Rastall, Rastall1}. In this theory the energy-momentum tensor is not a conserved quantity. The Rastall hypothesis is
\begin{equation}
 \nabla^{\mu}T_{\mu\nu} = \left(\frac{\kappa}{16\pi}\right)\nabla_{\nu}R,
\end{equation}
where $\kappa$ is a constant and $R$ is the Ricci scalar. However the Bianchi identity is maintained, i.e.,
\bea
\nabla^\nu G_{\mu\nu}=0,\label{Bian}
\eea
with  $G_{\mu\nu}\equiv R_{\mu\nu}-\frac{1}{2}Rg_{\mu\nu}$ being the Einstein tensor. An important note, the Rastall modification takes place only in the matter part of the theory and it leaves the geometric part of the field equations unchanged. This hypothesis is based on the fact that the non-zero divergence of the energy-momentum tensor has not yet been ruled out experimentally, then it can be questionable at least in a curved space-time. Then,  Rastall field equations write as
\begin{equation}
 R_{\mu\nu} -\frac{1}{2}g_{\mu\nu}R = 8\pi\left[T_{\mu\nu} -\left(\frac{\gamma - 1 }{2}\right)g_{\mu\nu}T\right],\label{eq4}
\end{equation}
where $T=g^{\mu\nu}T_{\mu\nu}$ is the trace of energy-momentum tensor and $\gamma$ is a dimensionless constant connected to $\kappa$. These field equations reduce to general relativity field equations in the limit of $\gamma=1$.

Since the Bianchi identities are still valid, when the divergence of eq. (\ref{eq4}) is taken the divergence of the energy-momentum tensor becomes
\begin{equation}
  \nabla^{\mu}T_{\mu\nu} = \left(\frac{\gamma - 1 }{2}\right)\nabla_{\nu}T.\label{FR}
\end{equation}
Thus the relation between $\kappa$ and $\gamma$ is
\begin{equation}
 \gamma = \frac{1-3\kappa}{1-2\kappa}.
\end{equation}

In order to obtain a divergence-free energy-momentum tensor, a new tensor is defined as
\begin{equation}
\tilde{T}_{\mu\nu} = T_{\mu\nu}  - \left(\frac{\gamma - 1 }{2}\right)g_{\mu\nu}T,\label{REMT}
\end{equation}
such that the field equations become
\begin{equation}
 G_{\mu\nu} = 8\pi \,\tilde{T}_{\mu\nu},\label{8}
\end{equation}
with $ \tilde{T}_{\mu\nu}$ being a quantity covariantly conserved.

One criticism of the Rastall gravity theory is that it has not a Lagrangian formulation widely accepted and complete. In the next section two different approaches are discussed as a possible Lagrangian formulation to obtain the Rastall gravity.

\section{Lagrangian formulation for Rastall gravity}

The field equations of the Rastall gravity theory were obtained in an \textit{ad hoc} way by Rastall, i.e., these equations do not come from a variational principle. In this section a Lagrangian formalism to this theory is discussed. Two theories of gravitation are presented that lead to Rastall gravity as a particular case. These modified gravity theories have as an important feature, the energy-momentum tensor of matter may be not covariantly conserved.

\subsection{$F(R,T)$ gravity}

The action that describes the $F(R,T)$ gravity model \cite{Harko} is
\begin{equation}\label{acao}
 S = \frac{1}{16 \pi }\int \sqrt{-g}\left[F(R,T)+16\pi \, {\cal L}_{m}\right]d^4x,
\end{equation}
where $f(R,T)$ is a function of the Ricci scalar $R$ and of the trace of the energy-momentum tensor $T=g^{\mu\nu}T_{\mu\nu}$ and ${\cal L}_m$ is the matter Lagrangian density.

In order to obtain the field equations, the variational principle is used, i.e.,  $\delta S = 0$. By varying the action $S$ with respect to the metric tensor components $g^{\mu\nu}$ we have
\begin{equation}
 \delta S = \frac{1}{16 \pi }\int\delta \{ \sqrt{-g}\left(F(R,T)+16\pi  {\cal L}_{m}\right)\} d^4x = 0,
\end{equation}
where
\begin{eqnarray}
 \delta [\sqrt{-g}F(R,T)] &=&  \delta(\sqrt{-g}) F(R,T) + \sqrt{-g}\delta (F(R,T)) \label{eq11}\\
 \delta (\sqrt{-g}{\cal L}_m) &=&  \delta(\sqrt{-g}){\cal L}_m + \sqrt{-g}\delta( {\cal L}_m).
 \end{eqnarray}
 Using that
 \bea 
 \delta \sqrt{-g} = -\frac{1}{2}\sqrt{-g}g_{\mu\nu}\delta g^{\mu\nu},
 \eea
the eq. (\ref{eq11}) becomes
\begin{equation}
 \delta[\sqrt{-g}F(R,T)] = -\frac{1}{2}\sqrt{-g}g_{\mu\nu}F(R,T)\delta g^{\mu\nu} + \sqrt{-g}\delta F(R,T)
\end{equation}
where
 \begin{equation}
 \delta F(R,T) = F_R\delta R + F_T\delta T
  \end{equation}
with $F_R = \partial F(R,T)/\partial R$  and  $F_T = \partial F(R,T)/\partial T$. Using that $R=g^{\mu\nu}R_{\mu\nu}$ we get
\begin{equation}
 \delta F(R,T) = F_R(R_{\mu\nu}\delta g^{\mu\nu}+ g^{\mu\nu}\delta R_{\mu\nu}) + F_T\delta T.
\end{equation}
By varying the trace of the energy-momentum tensor with respect to the metric tensor we obtain
\begin{eqnarray}
\frac{\delta T}{\delta g^{\mu\nu}} &=& \frac{\delta (g^{\alpha\beta}T_{\alpha\beta})}{\delta g^{\mu\nu}} =  \frac{\delta g^{\alpha\beta}}{\delta g^{\mu\nu}}T_{\alpha\beta} + g^{\alpha\beta}\frac{\delta T_{\alpha\beta}}{\delta g^{\mu\nu}} \nonumber \\ 
&=&  T_{\mu\nu} + \Theta_{\mu\nu},
\end{eqnarray}
where
\begin{equation}
\Theta_{\mu\nu} \equiv g^{\alpha\beta}\frac{\delta T_{\alpha\beta}}{\delta g^{\mu\nu}}.
\end{equation}
Considering that the energy-momentum tensor of the matter is defined as \cite{Landau}
\begin{equation}
 T_{\mu\nu} = -\frac{2}{\sqrt{-g}}  \frac{\delta \left(\sqrt{-g}{\cal L}_m\right)}{\delta g^{\mu\nu}}, \label{TEM}
\end{equation}
the variation of the action with respect to the metric tensor becomes
\begin{eqnarray}
 \delta S &=& \frac{1}{16 \pi }\int  \left[F_R R_{\mu\nu} - \frac{1}{2}g_{\mu\nu}F(R,T) + \left( T_{\mu\nu} + \Theta_{\mu\nu} \right)F_T - 8\pi \,T_{\mu\nu}\right. + \nonumber \\ 
 &{}&+ \left.g^{\alpha\beta}\frac{\delta R_{\alpha\beta}}{\delta g^{\mu\nu}}F_R\right]\delta g^{\mu\nu}\, \sqrt{-g}\,d^4x.
\end{eqnarray}

Using the Ricci tensor definition,
\begin{equation}
 R_{\mu\nu} = R^{\rho}\,_{\mu\rho\nu} = \partial_{\rho}\Gamma^{\rho}_{\nu\mu} - \partial_{\nu}\Gamma^{\rho}_{\rho\mu}  + \Gamma^{\lambda}_{\rho\lambda}\Gamma^{\rho}_{\nu\mu} - \Gamma^{\rho}_{\nu\lambda}\Gamma^{\lambda}_{\rho\mu}\,,
\end{equation}
where $\Gamma^{\rho}_{\nu\mu}$ is the Christoffel symbols, the variation of this tensor yields
\begin{equation}
   \delta R_{\mu\nu} = \nabla_{\rho}(\delta \Gamma^{\rho}_{\nu\mu})-  \nabla_{\nu}(\delta \Gamma^{\rho}_{\rho\mu}).\label{ricci}
 \end{equation}
Then
\begin{eqnarray}
  \delta S = \frac{1}{16 \pi }\int  d^4 x \sqrt{-g}
   \Bigl\{ \delta g^{\mu\nu} \Bigl [ \left(F_R R_{\mu\nu} - \frac{g_{\mu\nu}}{2}F(R,T)\right)  + \nonumber \\  + \left[\left( T_{\mu\nu} + \Theta_{\mu\nu} \right)F_T - 8\pi  T_{\mu\nu}\right]\Bigl ] 
  + g^{\alpha\beta}\left[\nabla_{\rho}\left( \delta\Gamma^{\rho}_{\beta\alpha}\right)  -  \nabla_{\beta}\left(\delta \Gamma^{\rho}_{\rho\alpha}\right)\right]F_R\Bigl\}. \label{eqCS}
\end{eqnarray}


The first and second parts of eq. (\ref{eqCS}) are known. Then let us calculate the third part separately, which is defined as
\begin{equation}
 I = \int \sqrt{-g}\left\{ g^{\alpha\beta}\left[\nabla_{\rho}\left( \delta\Gamma^{\rho}_{\beta\alpha}\right)  -  \nabla_{\beta}\left(\delta \Gamma^{\rho}_{\rho\alpha}\right)\right]F_R\right\} d^4x.
\end{equation}
For simplicity, let us write
\bea
I = I_1 + I_2,
\eea
where
\begin{eqnarray}
 I_1 &=& \int \sqrt{-g}\, g^{\alpha\beta}\nabla_{\rho}\left( \delta\Gamma^{\rho}_{\beta\alpha}\right) F_R\, d^4x  \label{eq26}\\
 I_2 &=& -  \int \sqrt{-g} \,g^{\alpha\beta}\nabla_{\beta}\left(\delta \Gamma^{\rho}_{\rho\alpha}\right)F_R\, d^4x. \label{eq27}
\end{eqnarray}
Let us calculate these terms separately. Using that
\begin{eqnarray}
 \sqrt{-g} g^{\alpha\beta}\nabla_{\rho}\left( \delta\Gamma^{\rho}_{\beta\alpha}\right) F_R =  \nabla_{\rho}\left(\sqrt{-g} g^{\alpha\beta} \delta\Gamma^{\rho}_{\beta\alpha}F_R\right)  - \sqrt{-g} g^{\alpha\beta}\nabla_{\rho}\left(F_R \right)\delta\Gamma^{\rho}_{\beta\alpha}\,,
\end{eqnarray}
eq. (\ref{eq26}) becomes
\begin{eqnarray}
I_1  &=&  \int  \nabla_{\rho}\left(\sqrt{-g} g^{\alpha\beta} \delta\Gamma^{\rho}_{\beta\alpha}F_R\right)d^4x  - \int \sqrt{-g} g^{\alpha\beta}\nabla_{\rho}\left(F_R \right)\delta\Gamma^{\rho}_{\beta\alpha}   \, d^4x.
\end{eqnarray}
Using the Gauss theorem, the first term may be written as a surface integral, i.e.,
\begin{equation}
 \int_R  \nabla_{\rho}\left(\sqrt{-g} g^{\alpha\beta} \delta\Gamma^{\rho}_{\beta\alpha}F_R\right)d^4x =  \int_{\partial R}  n_{\rho}V^{\rho}\, d \sigma 
\end{equation}
where $V^{\rho} = \sqrt{-g} g^{\alpha\beta} \delta\Gamma^{\rho}_{\beta\alpha}F_R$ and $n_{\rho}$ is normal to $\partial R$. Note that, $V^{\rho}$ is a vector field over a region $R$ with boundary $\partial R$. By taking that this surface integral is zero, we get
\begin{eqnarray}
 I_1 &=& - \int \sqrt{-g} g^{\alpha\beta}\nabla_{\rho}\left(F_R \right)\delta\Gamma^{\rho}_{\beta\alpha}   \, d^4x.\label{I1}
\end{eqnarray}
Similarly, the term $I_2$ becomes
\begin{eqnarray}
 I_2 &=&  \int \sqrt{-g} g^{\alpha\beta}\nabla_{\beta}\left(F_R \right)\delta\Gamma^{\rho}_{\rho\alpha}   \, d^4x.\label{I2}
\end{eqnarray}

The variation of the Christoffel symbols is given by
\begin{equation}
\delta \Gamma^{\lambda}_{\ \mu\nu} = \frac{g^{\lambda\rho}}{2}\left(-\nabla_{\rho}\delta g_{\mu\nu} 
+\nabla_{\mu}\delta g_{\nu\rho} + \nabla_{\nu}\delta g_{\rho\mu}\right)\label{chrisT}
\end{equation}
and
\begin{equation}
 \delta \Gamma^{\lambda}_{\ \lambda\nu} = \frac{1}{2}g^{\lambda\rho}\nabla_{\nu}\delta g_{\lambda\rho}.
\end{equation}
Then eq. (\ref{I2}) becomes
\begin{equation}
  I_2 =  \int \sqrt{-g} g^{\alpha\beta}\nabla_{\beta}\left(F_R \right)\delta\Gamma^{\rho}_{\rho\alpha}   \, d^4x =\frac{1}{2} \int \sqrt{-g} g^{\alpha\beta}\nabla_{\beta}\left(F_R \right)g^{\lambda\rho}\nabla_{\alpha}\delta g_{\lambda\rho} d^4x.
\end{equation}
Considering that $0=\delta(g_{\lambda\rho}g^{\lambda\rho})=\delta(g_{\lambda\rho}) g^{\lambda\rho} +  g_{\lambda\rho}\delta(g^{\lambda\rho})$ and the metric compatibility, i.e., $\nabla_{\mu}g^{\mu\nu} =\nabla_{\rho}g^{\mu\nu}= 0$, we get
\begin{equation}
  I_2 =  -\frac{1}{2} \int \sqrt{-g} g^{\alpha\beta}g_{\lambda\rho}\nabla_{\beta}F_R  \nabla_{\alpha}\delta g^{\lambda\rho}\, d^4x.
\end{equation}
Using that
\begin{eqnarray}
g^{\alpha\beta}g_{\lambda\rho}\nabla_{\beta}F_R  \nabla_{\alpha}\delta g^{\lambda\rho} &=& \nabla_{\alpha}\left[g^{\alpha\beta}g_{\lambda\rho}\nabla_{\beta}F_R \delta g^{\lambda\rho}\right] -  g^{\alpha\beta}g_{\lambda\rho}\nabla_{\alpha}\nabla_{\beta}F_R \delta g^{\lambda\rho}
\end{eqnarray}
the expression $I_2$ is given as
\begin{eqnarray}
   I_2 =  -\frac{1}{2} \int \sqrt{-g} \left\{\nabla_{\alpha}\left[g^{\alpha\beta}g_{\lambda\rho}\nabla_{\beta}F_R \delta g^{\lambda\rho}\right] -  g^{\alpha\beta}g_{\lambda\rho}\nabla_{\alpha}\nabla_{\beta}F_R \delta g^{\lambda\rho}\right\} \, d^4x.
\end{eqnarray}
Applying the Gauss theorem in the first term, it is transformed in a surface integral which is taken as zero. Then
\begin{eqnarray}
    I_2 =\frac{1}{2} \int \sqrt{-g}  g^{\alpha\beta}g_{\lambda\rho}\nabla_{\alpha}\nabla_{\beta}\, F_R \, \delta g^{\lambda\rho} \, d^4x.\label{I22}
\end{eqnarray}

Now, back to the term $I_1$ and using eq. (\ref{chrisT}) we get
\begin{eqnarray}
  I_1 &=& - \int \sqrt{-g}\, \nabla_{\rho}\left(F_R \right) g^{\alpha\beta} g^{\rho\lambda} \nabla_{\beta}\delta g_{\alpha\lambda}\, d^4x   -\frac{1}{2}\int \sqrt{-g}\,g_{\alpha\beta}\, g^{\rho\lambda}\nabla_{\lambda}\delta g^{\beta\alpha}\, d^4x.
\end{eqnarray}
 
The variational of  $\delta (g^{\alpha\beta} g^{\rho\lambda}  g_{\alpha\lambda})$ leads to $g^{\alpha\beta} g^{\rho\lambda}  \delta g_{\alpha\lambda}=-\delta g^{\beta\rho}$, then
\begin{eqnarray}
   I_1 &=& \int \sqrt{-g}\, \nabla_{\rho}\left(F_R \right) \nabla_{\beta} \, \delta g^{\rho\beta}\, d^4x   -\frac{1}{2}\int \sqrt{-g}\,g_{\alpha\beta}\, g^{\rho\lambda}\nabla_{\lambda}\delta g^{\beta\alpha}\, d^4x.\label{last}
\end{eqnarray}
Using similar steps as for the expression $I_2$, and the Gauss theorem, the eq. (\ref{last}) becomes
\begin{eqnarray}
   I_1 &=&  -  \int \sqrt{-g}\,
 \nabla_{\beta} \nabla_{\rho}\left(F_R \right)  \, \delta g^{\rho\beta}\, d^4x  +\frac{1}{2} \int \sqrt{-g}  g^{\alpha\beta}g_{\lambda\rho}\nabla_{\alpha}\nabla_{\beta}\, F_R \, \delta g^{\lambda\rho} \, d^4x. \label{I11}
\end{eqnarray}

With eqs. (\ref{I22}) and (\ref{I11}) the third term of eq. (\ref{eqCS}) is written as
\begin{eqnarray}
 I &=& \int \sqrt{-g}\left[ g^{\alpha\beta}g_{\mu\nu}\nabla_{\alpha}\nabla_{\beta}\, F_R - \nabla_{\mu} \nabla_{\nu}\,F_R \,\right]\delta g^{\mu\nu} \, d^4x \nonumber \\
 &=&\int \sqrt{-g}\left[ \left(g_{\mu\nu}\Box\,  - \nabla_{\mu} \nabla_{\nu}\right)\,F_R\right]\delta g^{\mu\nu} \, d^4x,
\end{eqnarray}
where $\Box=\nabla_\mu\nabla^\mu$ is the covariant d'Alembertian operator. Therefore, from eq. (\ref{eqCS}) the field equations are obtained as
\begin{equation}
     R_{\mu\nu}F_R - \frac{g_{\mu\nu}}{2} F(R,T)  + \left( T_{\mu\nu} + \Theta_{\mu\nu} \right)F_T   + (g_{\mu\nu} \Box - \nabla_{\mu}\nabla_{\nu})F_R  = 8\pi T_{\mu\nu}.\label{EqGen}
\end{equation}
By taking $F(R,T) = F(R) = R$ the general relativity is recovered.


By contracting eq. (\ref{EqGen}), a relation between the Ricci scalar and the trace of the energy-momentum tensor is obtained as
\begin{equation}
 R  F_R + 3\Box F_R - 2 F(R,T)  = 8\pi  T - (T + \theta)  F_T,   
\end{equation}
where $T = g^{\mu\nu}T_{\mu\nu}$ e $\theta = g^{\mu\nu}\Theta_{\mu\nu}$.
 
It is important to note that, the Rastall gravity theory is obtained for the particular case
  \begin{equation}
F(R,T) =  (1-\alpha)R + 8\pi  \,\alpha\beta\, T,\label{Escolha}
\end{equation}
where $\alpha$ and $\beta$ are free parameters. Using eq. (\ref{Escolha}) the field equations become
\begin{equation}
     R_{\mu\nu} - \frac{1}{2}g_{\mu\nu}R    =\frac{8 \pi  }{1-\alpha}\left[(1-\alpha\beta) T_{\mu\nu} +  \alpha\beta \left( \frac{g_{\mu\nu}}{2} T  - \Theta_{\mu\nu} \right)\right].
\end{equation}
By taking the divergence of this equation and using Bianchi identity (\ref{Bian}) we get
\begin{equation}
\nabla^{\mu}T_{\mu\nu}  = \frac{\alpha\beta}{2(\alpha\beta - 1)} \nabla_{\nu}T^* ,
\end{equation}
where $T^*\equiv g^{\mu\nu}\left(g_{\mu\nu}T-2\Theta_{\mu\nu}\right)$.
Choosing the free parameters as
\begin{eqnarray}
\beta = 1 \quad\quad\quad \rm{and} \quad\quad\quad \alpha = \frac{\gamma - 1 }{\gamma - 2},
\end{eqnarray}
the Rastall proposal is explicitly obtained as
\begin{equation}
\nabla^{\mu}T_{\mu\nu}  = \frac{\gamma - 1}{2} \nabla_{\nu}T^*.\label{Rastall_Const}
\end{equation}
Note that, when $\gamma=1$ the general relativity is recovered. Therefore the Rastall theory of gravity may be constructed as a particular case of the $F(R,T)$ gravity. An important note, to be rigorous,  eq. (\ref{Rastall_Const}) is a Rastall-type condition, since $\nabla^\mu\Theta_{\mu\nu}$ is not zero. However, the eqs. (\ref{Rastall_Const}) and (\ref{FR}) exhibit the same interpretation, i.e., the non-conservation of the energy-momentum tensor.

\subsection{$F(R,{\cal L}_{m})$ gravity}

Another proposal to obtain the Rastall gravity is through of the field equations of the $F(R,{\cal L}_{m})$ gravity. The action for this modified theory of gravity takes the following form 
\begin{equation}\label{acaofrlm}
S = \frac{1}{16 \pi }\int \sqrt{-g}\left[ F(R,{\cal L}_{m}) \right] d^4x,
\end{equation}
where $F(R,{\cal L}_{m})$ is an arbitrary function of the Ricci scalar $R$, and of the Lagrangian density corresponding to matter source, ${\cal L}_m$. 

Let's assume that the Lagrangian density ${\cal L}_m$ of the matter depends only on the metric tensor components, and not on its derivatives. Then the energy-momentum tensor defined as in eq. (\ref{TEM}) is given as
\begin{equation}\label{tensoreneg2}
T_{\mu\nu} = g_{\mu\nu} {\cal L}_{m} - 2 \frac{\partial {\cal L}_{m}}{\partial g^{\mu\nu}}.
\end{equation}

By varying the action (\ref{acaofrlm}) with respect to the metric tensor component we obtain
\begin{equation}
\delta S = \frac{1}{16 \pi }\int \sqrt{-g}\left[F_R \delta R + F_{({\cal L}_{m})}\frac{\delta {\cal L}_{m}}{\delta g^{\mu\nu}}\delta g^{\mu\nu} - \frac{1}{2}g_{\mu\nu}F(R,{\cal L}_{m})\delta g^{\mu\nu}\right]d^4x,
\end{equation}
where $F_{({\cal L}_{m})} =\partial F(R,{\cal L}_{m})/\partial {\cal L}_{m}$. Using the variations of the Ricci tensor (\ref{ricci}) and Christoffel symbols (\ref{chrisT}), the variation of the action of the gravitational field becomes
\begin{eqnarray}
\delta S &=&  \frac{1}{16 \pi }\int \sqrt{-g}\left[F_R R_{\mu\nu}\delta g^{\mu\nu} +
F_R g_{\mu\nu}\nabla_{\mu}\nabla^{\mu} \delta g^{\mu\nu} - F_R\nabla_{\mu}\nabla_{\nu}\delta g^{\mu\nu} + \right.\nonumber \\
&{}&\left.+ F_{({\cal L}_{m})}\frac{\delta {\cal L}_{m}}{\delta g^{\mu\nu}}\delta g^{\mu\nu} - \frac{1}{2}g_{\mu\nu}F(R,{\cal L}_{m})\delta g^{\mu\nu}\right]d^4x.\label{eq37}
\end{eqnarray}
Following similar steps to those developed in the previous subsection, and with the use of the definition of the energy-momentum tensor, eq. (\ref{TEM}), the field equations are
\begin{equation}\label{campoflm}
F_R R_{\mu\nu} + (g_{\mu\nu}\Box - \nabla_{\mu}\nabla_{\nu})F_R - \frac{1}{2}\left[F(R,{\cal L}_{m}) - F_{({\cal L}_{m})}{\cal L}_{m}\right]g_{\mu\nu} = \frac{1}{2}F_{({\cal L}_{m})}T_{\mu\nu} .
\end{equation}

By taking a particular case, such as
\begin{equation}
F(R,{\cal L}_{m}) = \alpha R + {\cal G}({\cal L}_{m})
\end{equation}
where $\alpha$ is an arbitrary constant to be determined and ${\cal G}({\cal L}_{m})$ is a function of the matter Lagrangian, the field equations (\ref{campoflm}) become
\begin{eqnarray}
 R_{\mu\nu} - \frac{1}{2}g_{\mu\nu} R  = \frac{1}{2\alpha}\left\{{\cal G}'T_{\mu\nu} + ({\cal G}   - {\cal G}' {\cal L}_{m}) g_{\mu\nu}\right\}
\end{eqnarray}
with ${\cal G}'$ being the derivative of ${\cal G}({\cal L}_{m})$ with respect the matter Lagrangian ${\cal L}_{m}$.

Aplying the divergent $\nabla^{\mu}$ and using the Bianchi identity, we obtain
\begin{eqnarray}
\frac{1}{2\alpha}\nabla^{\mu}\left\{ {\cal G}'T_{\mu\nu} + ({\cal G}   - {\cal G}' {\cal L}_{m}) g_{\mu\nu}\right\}=0.
\end{eqnarray}
This result leads to the Rastall gravity if
\bea
\nabla^{\mu}T^{(eff)}_{\mu\nu}=\left(\frac{\gamma -1 }{2}\right)\nabla_{\nu}T
\eea
where
\begin{equation}
T^{(eff)}_{\mu\nu} \equiv \frac{1}{2\alpha}\left\{ {\cal G}'T_{\mu\nu} + ({\cal G}   - {\cal G}' {\cal L}_{m}) g_{\mu\nu}\right\}.
\end{equation}
It is important to note that, the Rastall-type condition is satisfied by any ${\cal G}( {\cal L}_{m})$ function. 

The results obtained here show that the Rastall theory of gravity emerges as a particular case from modified gravity theories. Therefore, its field equations have basis on a variational principle. In the next section, let's use the Rastall gravity to study the causality issue in the G\"{o}del-type universe.

\section{G\"{o}del-type universe}

Here the main characteristics of the G\"{o}del-type metric are presented. In cylindrical coordinates the G\"{o}del-type metric is given by
\begin{equation}
 ds^2 = \left[dt+H(r)d\phi\right]^2 - D(r)d\phi^2 - dr^2 -dz^2
\end{equation}
where
\bea
H(r)&=&\frac{4\Omega}{m^2}\sinh^2\left(\frac{mr}{2}\right)\\
D(r)&=&\frac{1}{m}\sinh(mr)
\eea
with $\omega$ and $m$ being free parameters. The G\"{o}del solution is a particular case in which $m^2 = 2\Omega^2$.

In order to analyze causal and non-causal regions, the G\"{o}del-type line-element is written as
\begin{equation}
 ds^2 = dt^2 + 2H(r)d\phi dt - \mathbb{G}(r)d\phi^2 - dr^2 -dz^2
\end{equation}
where
\begin{eqnarray}
 \mathbb{G}(r) &=& D^2(r) - H^2(r) \\
 &=& \frac{4}{m^2} \sinh\left(\frac{mr}{2}\right)\left[1+ \left(1-\frac{4\Omega^2}{m^2}\right)\sinh^2\left(\frac{mr}{2}\right)\right].
\end{eqnarray}
The existence of CTC's in G\"{o}del-type universe, i.e. circles defined by $t, z, r = \rm{const}$, depend on the behavior of the function $ \mathbb{G}(r) $. The case where $\mathbb{G}(r)<0$ leads to the existence of CTC's, that is determined by circles defined by $t, z, r = \rm{const}$. The non-causal regions occur for $r > r_c$ such that
\begin{equation}
 \sinh^2\left(\frac{mr_c}{2}\right) = \left(\frac{4\Omega^2}{m^2}-1\right)^{-1},\label{CR}
\end{equation}
where $r_c$ is the critical radius. Then the relation between the parameters $m$ and $\Omega$ determines causal and non-causal regions. For $0<m^2<4\Omega^2$ CTC's appear for $r> r_c$. If $m^2 =  4\Omega^2$ the critical radius goes to infinity, and then for $m^2 \geq 4\Omega^2$ there are no CTC's, and hence the breakdown of causality is avoided.

The line-element may be write as $ds^2 = g_{\mu\nu}dx^{\mu}dx^{\nu}$, where the metric tensor and its inverse are 
\begin{equation}
 g_{\mu\nu} =\begin{pmatrix}
                     1 & 0 & H & 0 \\
                     0 & -1 & 0 & 0 \\
                     H & 0 & -\mathbb{G} & 0 \\
                     0 & 0 & 0 & -1
                     \end{pmatrix}    ,                 
                     \quad\quad\quad
                      g^{\mu\nu} = \dfrac{1}{D^2}\begin{pmatrix}
                     \mathbb{G} & 0 & H& 0 \\
                     0 & -D^2 & 0 & 0 \\
                    H & 0 & -1& 0 \\
                     0 & 0 & 0 & -D^2
                     \end{pmatrix}.
\end{equation}

A further simplification comes about by the following choice of basis such as
\begin{equation}
 ds^2 = \eta_{ab}\Upsilon^{a}\Upsilon^{b}
\end{equation}
where  $\Upsilon^{a} = e^{a}_{\ \ \mu}dx^{\mu}$. Then
\begin{eqnarray}
\Upsilon^{(0)} = dt + H(r) d\phi &{ }& \Upsilon^{(1)} = dr \\
 \Upsilon^{(2)} = \mathbb{G}(r) d\phi &{ }& \Upsilon^{(3)} = dz 
\end{eqnarray}
with the tetrads given by
\begin{equation}
 e_{a\mu} =\begin{pmatrix}
                     1 & 0 & H & 0 \\
                     0 & -1 & 0 & 0 \\
                     0 & 0 & -D & 0 \\
                     0 & 0 & 0 & -1
                     \end{pmatrix}
                     \quad\quad \mathrm{and} \quad\quad
                     e^{a}_{\ \ \mu} =\begin{pmatrix}
                     1 & 0 & H & 0 \\
                     0 & 1 & 0 & 0 \\
                     0 & 0 & D & 0 \\
                     0 & 0 & 0 & 1
                     \end{pmatrix},
\end{equation}
where $ e^{a}_{\ \ \mu} = \eta^{ab}e_{b\mu}$ has been used. Note that, the tetrads can be obtained through the normalization, i.e.,  $e^{a}_{\ \ \mu}e_{a\nu} = g_{\mu\nu}$ and $e_{a}^{\ \ \mu}e_{b\mu} = \eta_{ab}$.

In the next section, the field equations of Rastall gravity will be analyzed in a local space-time for different matter contents.

\section{G\"{o}del-type solution in Rastall gravity}

Here our main objective is to solve the Rastall equations (\ref{8}) in a local space-time. These equations are given as
\begin{equation}
 G_{ab} = 8\pi \tilde{T}_{ab},\label{FE}
\end{equation}
where $\tilde{T}_{ab}$ is defined in eq. (\ref{REMT}) and the Einstein tensor is
\begin{equation}
 G_{ab} = R_{ab} - \frac{1}{2}\eta_{ab}R.
\end{equation}
The non-vanishing components of the Einstein tensor $G_{ab}$ take the form
\begin{eqnarray}
 G_{(0)(0)} &=& 3\Omega^2 - m^2 \nonumber \\
 G_{(1)(1)} &=& G_{(2)(2)} = \Omega^2 \nonumber\\
 G_{(3)(3)} &=& m^2 - \Omega^2. 
\end{eqnarray}

An important ingredient to study the causality problem is the matter source. Then let's solve the field equations for two different matter sources: (i) perfect fluid and (ii) perfect fluid plus scalar field.

\subsection{Non-causal solution}

Here a perfect fluid of density $\epsilon$ and pressure $p$ is considered. Its energy-momentum tensor is defined as
\begin{equation}
 T_{ab} = (\epsilon +p) u_{a}u_{b} + (\Lambda'-p)\eta_{ab},
\end{equation}
where $\Lambda' = \Lambda/8\pi$ with $\Lambda$ as the cosmological constant and $u_{a}$ is the four-velocity of the fluid. In addition, it is known that $u^{\mu} = e_{(0)}^{\ \ \mu}$. Then 
\bea
 u^{a} = e^{a}_{\ \ \mu}u^{\mu} = e^{a}_{\ \ \mu}e_{(0)}^{\ \ \mu} = e^{a}_{\ \ 0}.
\eea
Therefore the four-velocity in a local and non-local space-time is
\bea
u^a = (1,0,0,0) \quad\quad \mathrm{and} \quad\quad  u^{\mu} = \left(1,0,H(r),0\right).
\eea

Then all components of the energy-momentum tensor of the perfect fluid is written as
\begin{equation}
T_{ab} =\begin{pmatrix}
                     \epsilon+\Lambda' & 0 & 0 & 0 \\
                     0 & p-\Lambda' & 0 & 0 \\
                     0 & 0 & p-\Lambda'& 0 \\
                     0 & 0 & 0 & p-\Lambda'
                     \end{pmatrix}.
\end{equation}
The trace of the energy-momentum tensor is
\begin{eqnarray}
 T \equiv \eta^{ab}T_{ab}=\epsilon + 4\Lambda' - 3p.
\end{eqnarray}
Using these results, the components of the energy-momentum tensor due to the Rastall gravity, defined in eq.(\ref{REMT}), become
\begin{eqnarray}
 \tilde{T}_{(0)(0)} &=& \left(\epsilon + \Lambda' \right) - \left(\frac{\gamma - 1}{2}\right)\left[\epsilon + 4\Lambda'-3p\right]  \nonumber \\
 \tilde{T}_{(1)(1)} &=&\tilde{T}_{(2)(2)}=\tilde{T}_{(3)(3)}= \left(p - \Lambda' \right) + \left(\frac{\gamma - 1}{2}\right)\left[\epsilon + 4\Lambda'-3p\right] .
\end{eqnarray}

For this matter source, the field equations (\ref{FE}) are
\begin{eqnarray}
 3\Omega^2 - m^2 &=& \left(\epsilon + \Lambda' \right) - \left(\frac{\gamma - 1}{2}\right)\left[\epsilon + 4\Lambda'-3p\right] \label{eq1} \\
 \Omega^2 &=& \left(p - \Lambda' \right) + \left(\frac{\gamma - 1}{2}\right)\left[\epsilon + 4\Lambda'-3p\right]   \label{eq2} \\
m^2-\Omega^2  &=& \left(p - \Lambda' \right) + \left(\frac{\gamma - 1}{2}\right)\left[\epsilon + 4\Lambda'-3p\right].  \label{eq3}
\end{eqnarray}
Equations (\ref{eq2}) and (\ref{eq3}) give
\begin{equation}
 m^2 = 2\Omega^2.
\end{equation}
This relation defines the G\"{o}del metric. The remaining field equations reduce to
\begin{eqnarray}
 \epsilon   &=& \frac{ p[4-3\gamma] - 2\Lambda'[3-2\gamma]}{[2-\gamma]}  \\
 m^2        &=& \frac{2p[3-2\gamma] - 2\Lambda'[3-2\gamma]}{[2-\gamma]}.
\end{eqnarray}
Then in the framework of Rastall gravity the critical radius $r_c$ is given by
\bea
r_c=\frac{2}{m}\sinh^{-1}(1)=2\sinh^{-1}(1)\sqrt{\frac{2-\gamma}{(3-2\gamma)(2p-2\Lambda')}}.
\eea
Therefore, this result shows that beyond the critical radius exist non-causal G\"{o}del circles, which depends on the Rastall parameter and the matter content.

\subsection{Causal solution}

In the last subsection has been seen that the G\"{o}del solution in Rastall gravity with a perfect fluid as matter source leads to breakdown of causality. Now let's investigate whether causal solution emerges from other matter sources. Here a combination of a perfect fluid with a scalar field is considered as matter source. Its energy-momentum tensor is given by
\begin{equation}
 T^*_{\mu\nu} =  T_{\mu\nu} + T_{\mu\nu}^{\Phi}, \label{total}
\end{equation}
where $T_{\mu\nu}$ corresponds to a perfect fluid and $T_{\mu\nu}^{\Phi}$ describes a scalar field defined by
\begin{equation}
 T_{\mu\nu}^{\Phi} = \nabla_{\mu}\Phi\nabla_{\nu}\Phi -\frac{1}{2}g_{\mu\nu}g^{\rho\sigma}\nabla_{\rho}\Phi\nabla_{\sigma}\Phi.
\end{equation}
Using the local basis $\Upsilon^{a} = e^{a}_{\ \ \mu}dx^{\mu}$,  the energy-momentum tensor associated to scalar field becomes
\begin{equation}
  T_{ab}^{\Phi} = \nabla_{a}\Phi\nabla_{b}\Phi -\frac{1}{2}\eta_{ab}\eta^{cd}\nabla_{c}\Phi\nabla_{d}\Phi.
\end{equation}

By taking that the scalar field has the form $\Phi(z) = Az + \rm{cte}$, with $A$ being a constant, the non-zero components of energy-momentum tensor are
\begin{eqnarray}
  T_{(0)(0)}^{\Phi} &=&T_{(3)(3)}^{\Phi}= \frac{A^2}{2}\\
  T_{(1)(1)}^{\Phi} &=&T_{(2)(2)}^{\Phi}= -\frac{A^2}{2}.  
\end{eqnarray} 
Then the non-vanishing components of the total energy-momentum tensor,  eq. (\ref{total}), are
\begin{eqnarray}
 T^*_{(0)(0)} &=& \left(\epsilon + \Lambda' \right) + \frac{A^2}{2}\nonumber \\
 T^*_{(1)(1)} =T^*_{(2)(2)}&=& \left(p - \Lambda' \right) -\frac{A^2}{2}  \nonumber \\
 T^*_{(3)(3)} &=& \left(p - \Lambda' \right)+ \frac{A^2}{2},
\end{eqnarray}
and the trace of $T^*_{ab}$ is
\begin{equation}
 T^* = \epsilon+ 4\Lambda'- 3p +A^2. 
\end{equation}
In order to obtain the field equations, the non-null components of the tilde energy-momentum tensor, defined in eq. (\ref{REMT}), are written as
\begin{eqnarray}
 \tilde{ {T}}_{(0)(0)} &=& \left(\epsilon + \Lambda'+ \frac{A^2}{2} \right) - \left(\frac{\gamma - 1}{2}\right)\left[\epsilon + 4\Lambda'-3p+A^2\right]  \nonumber \\
  \tilde{ {T}}_{(1)(1)} &=& \left(p - \Lambda' -\frac{A^2}{2}\right) + \left(\frac{\gamma - 1}{2}\right)\left[\epsilon + 4\Lambda'-3p+A^2\right]  \nonumber \\
  \tilde{ {T}}_{(2)(2)} &=& \left(p - \Lambda' -\frac{A^2}{2}\right) + \left(\frac{\gamma - 1}{2}\right)\left[\epsilon + 4\Lambda'-3p+A^2\right]  \nonumber \\
  \tilde{ {T}}_{(3)(3)} &=& \left(p - \Lambda'+ \frac{A^2}{2} \right) + \left(\frac{\gamma - 1}{2}\right)\left[\epsilon + 4\Lambda'-3p+A^2\right]. 
\end{eqnarray}
Then the field equations for the combined matter source take the form
\begin{eqnarray}
 3\Omega^2 - m^2  &=& \left(\epsilon + \Lambda' + \frac{A^2}{2}\right) - \left(\frac{\gamma - 1}{2}\right)\left[\epsilon + 4\Lambda'-3p+A^2\right] \label{eq1r} \\
 \Omega^2         &=& \left(p - \Lambda' - \frac{A^2}{2}\right) + \left(\frac{\gamma - 1}{2}\right)\left[\epsilon + 4\Lambda'-3p+A^2\right]   \label{eq2r} \\
m^2-\Omega^2      &=& \left(p - \Lambda' + \frac{A^2}{2}\right) + \left(\frac{\gamma - 1}{2}\right)\left[\epsilon + 4\Lambda'-3p+A^2\right].  \label{eq3r}
\end{eqnarray}
These equations have as solution
\bea
m^2 - 2\Omega^2 &=& A^2\\
\rho&=&\frac{1}{2}\left(2\Omega^2-A^2\right),\\
p&=&\frac{A^2(\gamma-5)+4\Lambda'(2\gamma-3)+2\Omega^2(\gamma-1)}{6(\gamma-1)}.
\eea
By considering that these equations satisfy the condition $A^2>0$, a causal G\"{o}del-type class of solutions emerge when
\bea
m^2=4\Omega^2.\label{eq86}
\eea
This condition applied to eq. (\ref{CR}) leads to $r_c\rightarrow\infty$. Hence, there is no violation of causality of G\"{o}del-type in Rastall gravity for this combination of matter source. It is important to note that, if only a simple scalar field $\Phi(z)$ is considered as matter source the causal G\"{o}del-type solution persists in this gravitational theory.

\section{Conclusion}

The Rastall theory of gravity is a gravitational model that does not exhibit the usual conservation law of the energy-momentum tensor. In this theory a non-minimal coupling of matter fields to geometry is considered where the divergence of the energy-momentum tensor is proportional to the gradient of the Ricci scalar such that the usual conservation law is recovered in the flat space-time. Although numerous studies have been developed in this theory, it is highly criticized due to the lack of a Lagrangian formalism. Here the modified gravity theories $f(R,T)$ and $f(R, {\cal L}_m)$ have been used to construct a possible Lagrangian formalism to the Rastall theory. Then the G\"{o}del-type universe is studied in this gravitational model and the question of the causality violation is examined. By considering the perfect fluid as matter source is shown that the G\"{o}del solution is recovered. Thus the Rastall theory allows causality violation. An expression for the critical radius (beyond which the causality is violated) is determined, which depends on parameters of the gravity theory and the matter content. Also has been examined whether other matter sources could lead to the causal solution. Then a combination of perfect fluid
with a scalar field as a matter content has been considered and thus a causal solution is found such that the causality violation is not allowed. Therefore the Rastall gravity permits both causal and non-causal G\"{o}del-type solution. An important note, the analysis developed here is completely different of the analysis realized in \cite{Ulhoa}, since the G\"{o}del-type metrics, the results and discussions are different. Here the parameters of the metric allow a more complete discussion about causal and non-causal solutions.

\section*{Acknowledgments}

This work by A. F. S. is supported by CNPq projects 308611/2017-9 and 430194/2018-8; W. A. G. M. thanks CAPES for financial support.

\end{document}